\documentclass[showpacs,showkeys,aps,twocolumn,superscriptaddress,
longbibliography]{revtex4-1}
\usepackage[utf8]{inputenc}
\usepackage[T1]{fontenc}
\usepackage{xspace}
\usepackage[super]{nth}
\usepackage{multirow}
\usepackage[normalem]{ulem}
\usepackage{graphicx}
\usepackage{epstopdf}
\usepackage{color,xcolor}
\usepackage{epsfig}
\usepackage{epstopdf}
\usepackage{amsmath,amsfonts,amssymb}
\usepackage{bm}
\usepackage{nicefrac}
\usepackage{upgreek}
\usepackage[EULERGREEK]{sansmath}
\usepackage{physics}
\usepackage[acronym]{glossaries}
\usepackage[colorinlistoftodos,prependcaption,textsize=tiny]{todonotes}
\usepackage{siunitx}
\usepackage[
    colorlinks = true,
    citecolor  = blue,
    urlcolor   = blue,
  ]{hyperref}
\newcommand{\etal}{\textit{et~al.}\xspace}

\newcommand{\MnOI}{(Mn--O)$_{z}$\xspace}
\newcommand{\MnOIIa}{(Mn--O)$_{x1}$\xspace}
\newcommand{\MnOIIb}{(Mn--O)$_{x2}$\xspace}
\newcommand{\dOI}{\ensuremath{d_z}}      %d_z
\newcommand{\dOIIa}{\ensuremath{d_{x1}}} %d_{2_{\text{I}}}  %d_{x1}
\newcommand{\dOIIb}{\ensuremath{d_{x2}}} %d_{2_{\text{II}}} %d_{x2}
\newcommand{\eg}{\ensuremath{e_{\text{g}}}\xspace}

\newcommand{\tg}{\ensuremath{t_{2\text{g}}}\xspace}

\renewcommand{\vec}[1]{{{\bm #1}}}
\newcommand{\add}[1]{{\color{black}{#1}}}
\newcommand{\Fref}[2][]{Fig.~\ref{#2}\textcolor{black}{#1}} % beginning of sentences
\newcommand{\Tref}[1]{Table~\ref{#1}\xspace} % beginning of sentences
\newcommand{\sref}[1]{section~\ref{#1}\xspace}
\newcommand{\affJKU}{Institute for Theoretical Physics, Johannes Kepler University Linz, Altenberger Stra\ss{}e 69, 4040 Linz, Austria}
\newcommand{\affMLU}{Institute of Physics, Martin Luther University Halle-Wittenberg, Von-Seckendorff-Platz 1, 06120 Halle, Germany}

\newcommand{\affMPIhalle}{Max Planck Institute of Microstructure Physics, Weinberg 2, 06120 Halle, Germany}
\newcommand{\affUniTurku}{Department of Physics and Astronomy, University of Turku, FIN-20014 Turku, Finland}
\newcommand{\affMatSurf}{Turku University Centre for Materials and Surfaces (MatSurf), Turku, Finland}
\newcommand{\affWihuri}{Wihuri Physical Laboratory, Department of Physics and Astronomy, University of Turku, FI-20014 Turku, Finland}

\begin{document}
\title[First-principles investigation of the magnetic phase diagram 
of
Gd$_{1-x}$Ca$_{x}$MnO$_{3}$ ]{ First-principles investigations
of the magnetic phase diagram of Gd$_{1-x}$Ca$_{x}$MnO$_{3}$ }

\author{Hichem  \surname{Ben Hamed}}
\email{hichem.ben-hamed@physik.uni-halle.de}
\affiliation{\affMLU}

\author{Martin \surname{Hoffmann}}
\email{martin.hoffmann@jku.at}
\affiliation{\affJKU}

\author{Waheed A. \surname{Adeagbo}}
\affiliation{\affMLU}

\author{Arthur \surname{Ernst}}
\affiliation{\affJKU}
\affiliation{\affMPIhalle}

\author{Wolfram \surname{Hergert}}
\affiliation{\affMLU}

\author{Teemu Hynninen}
\affiliation{\affWihuri}

\author{Kalevi \surname{Kokko}}
\affiliation{\affUniTurku}
\affiliation{\affMatSurf}

\author{Petriina \surname{Paturi}}
\affiliation{\affWihuri}

\date{\today}
%%%%%%%%%%%%%%%%%%%%%%
\begin{abstract}
  We studied for the first time the magnetic phase diagram
  of the rare-earth manganites series Gd$_{1-x}$Ca$_{x}$MnO$_{3}$ (GCMO) over
  the full concentration range based on density functional theory. 
  GCMO has been shown to form solid solutions. We take into account 
  this disordered character
  by adapting special quasi random structures at different concentration
  steps. The magnetic phase diagram is mainly
  described by means of the magnetic exchange interactions between the Mn sites
  and Monte Carlo simulations were performed to estimate the corresponding
  transition temperatures. They agree very well with recent 
  experiments. The hole doped region $x<0.5$ shows a strong 
  ferromagnetic ground state, which competes with A-type
  antiferromagnetism at higher Ca concentrations $x>0.6$.
\end{abstract}
%%%%%%%%%%%%%%%%%%%%%
\maketitle

% FOOTnotes
% first footnote cite with Note1
\footnotetext{The KKR-CPA method is implemented in our code HUTSEPOT
  \cite{Geilhufe2015}. 
  Calculations for GCMO with this code are on the way, but not
  completed up to now.}
% APS style, PRB
% second footnote cite with Note2
\footnotetext{See
  Supplemental Material at [URL will be inserted by publisher] for
  details about the computational setup, validation
  of the choice of exchange-correlation functionals, magnetic moments,
  density of states, magnetic exchange parameters,
  and the special quasi random structures.}
% third footnote cite with Note3
\footnotetext{Experimental lattice constants were provided by Wihuri
  Physical Laboratory, University of Turku, Finland.}
% third footnote cite with Note4
\footnotetext{All ionic radii were taken from 
	\url{http://abulafia.mt.ic.ac.uk/shannon/}, which 
	is based on Ref. \cite{Shannon:a12967}.}
% ..........................................................
% SECTION ..................................................
% ..........................................................
\section{Introduction} % ...................................
\label{sec:introduction}
Transition metal oxides are of current interest and constitute one
class of promising materials to spawn diverse semiconductor
devices \cite{Mannhart1607}.
% surpass the silicon based technology\cite{Mannhart1607}.
They exhibit a wide range of exotic properties, owing mainly to the
partly filled \textit{d} shell \cite{Ismail-Beigi2017}.
The hybridization between oxygen $p$
states and the strongly correlated $3d$ states induce intriguing spin,
charge and orbital ordering. These properties are stimulated by the
close interplay of structural, electronic and magnetic degrees of
freedom. The discovery of the colossal magneto resistance (CMR)
effect \cite{ PhysRevLett.71.2331,RevModPhys.73.583} has triggered an
intensive study of the series of rare-earth manganese oxides with 
general formula $R$MnO$_{3}$ (with variable $R=\text{La}$, Ce, ...).

The $R$MnO$_{3}$ series consists of insulating
perovskites, which show a multitude of antiferromagnetic (AFM) structures
earlier studied by Kimura \etal \cite{PhysRevB.68.060403}. 
The observed A-type AFM (A-AFM) ground state was associated with the
tilting of the MnO$_{6}$ octahedron, known as GdFeO$_{3}$-type distortion.
This kind of distortion becomes even more pronounced for 
smaller ionic radius of the rare-earth ions ($r_{R}$).

Due to the perovskite structure of $R$MnO$_{3}$,
the resulting crystal-field breaks
the degeneracy of the Mn$^{3+}$ \textit{d} orbitals. %giving rise to 
Thus, they split into two degenerated orbitals ($\eg$)
  and three degenerated orbitals ($\tg$).
The strong Hund's coupling favors the parallel alignment
of the four electrons in the majority spin channel.
The
cooperative Jahn-Teller distortions lift in addition the double degeneracy of the
$\eg$ orbitals, while the $\tg$ orbitals become localized. 
The electrons occupying the
$\eg$ orbitals can in turn hop between the Mn sites through
the $p$ orbitals of oxygen. 
This mechanism %favors the metallic ferromagnetism. It 
is known
as the double exchange interaction mechanism and was earlier
introduced in the works of Zener \cite{PhysRev.82.403} and
Anderson \cite{PhysRev.100.675}.

Recently, a special focus on
$R$MnO$_{3}$ was raised because additional features can be 
accessed by modulating the electrical
charge carrier density. That can be realized, e.g., with applying an
electrostatic field \cite{0022-3727-38-8-R01} or chemical doping by introducing
alkaline earth elements (abbreviated as $A$) at the $R$ site. 

The incorporation of alkaline earth elements is the method
we want to focus on in this work because the $R$MnO$_{3}$
perovskite structure is very robust against adding other ions.
It is already widely used since the
early works of Wollan and Koehler \cite{PhysRev.100.545} and
Goodenough \cite{PhysRev.100.564}. 
Several material systems were already investigated and show 
full miscibility between the $R$ and $A$ elements, e.g.,
the La$_{1-x}$Ca$_{x}$MnO$_{3}$ series (LCMO)
\cite{PhysRev.100.545,PhysRev.100.564},
or the Pr$_{1-x}$Ca$_{x}$MnO$_{3}$ series (PCMO) \cite{PhysRevB.92.155148}.
In these solid solutions, the substitution of $R$
ions by $A$ ions causes the Mn $\eg$ electrons to hop to the neighboring ions
-- a four-valent Mn ion appears. Consequently, two types of manganese
emerge in the cell, namely, Mn$^{3+}$ and Mn$^{4+}$, and such systems are called
mixed valence manganites.

% ..........................................................
\begin{figure}
  \centering
  \includegraphics[width=246pt]{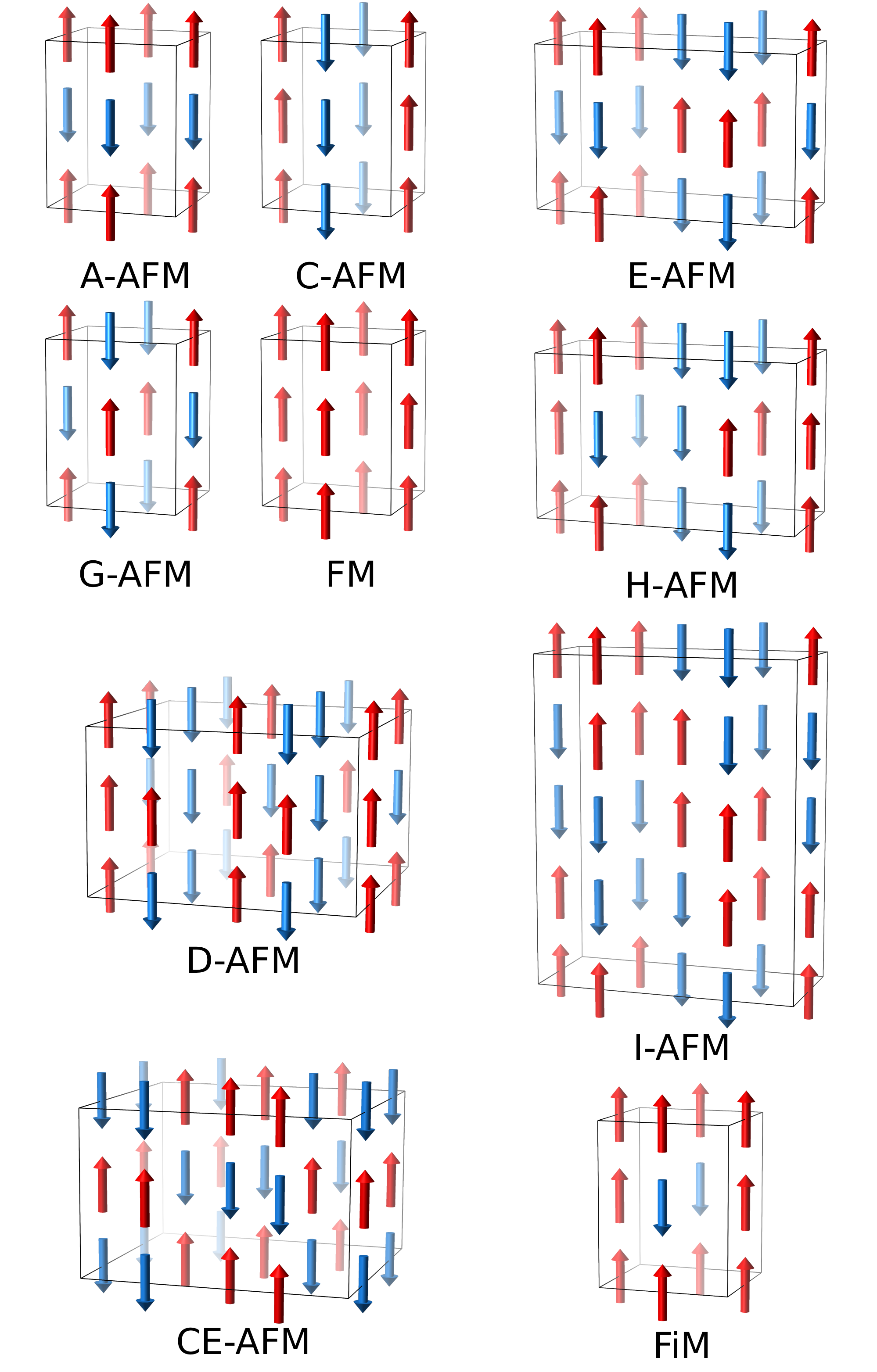}
  \caption{The different magnetic ground state structures
    which were suggested in \cite{PhysRev.100.545} and 
    were taken into account in this work.
    Here, only the magnetic moments
    at the Mn sites are represented as arrows -- red for 
    the majority and blue for the minority spin direction. 
    Different numbers of repeated Pbnm unit cells (see \Fref{fig:structure}) are
    needed to depict the antiferromagnetic (AFM) structures. The opacity of the 
    arrows has no particular meaning but only serves the perspective view.
    Structural figures were prepared with VESTA \cite{Momma2011jac}.
  }
  \label{fig:magnetic ground states}
\end{figure}
% ..........................................................

A prominent member of the $R$MnO$_{3}$ series is GdMnO$_3$.
The main reason is    
%Within this group of $R$MnO$_{3}$ perovskites, GdMnO$_3$ 
%could play an important role because of 
its location in the magnetoelectric phase diagram of 
the $R$MnO$_{3}$ compounds as a function of $r_{R}$:
in close vicinity of the 
collinear A-type AFM phase but also close to a ferroelectric state \cite{PhysRevB.71.224425, PhysRevLett.101.097204}. 
Hence, the phases could be manipulated rather easily
by external means. Kimura \etal \cite{PhysRevB.71.224425}
found, for instance, that a magnetic field of about \SI{1}{\tesla}
is sufficient to produce ferroelectricity.
On the other hand, GdMnO$_{3}$ could be also an
important candidate for future magneto-optic devices
because of its strong magneto–dielectric coupling \cite{Pimenov2006}.

Beiranvand \etal \cite{BEIRANVAND2017126} studied 
 the magnetic phase diagram of the Gd${_{1-x}}$Ca${_{x}}$MnO$_{3}$ series (GCMO)
using magnetoresistive measurements of magnetoresistivity in order 
to understand basic properties of this system.
They reported a rich and complicated magnetic phase diagram 
where the CMR effect showed up at doping 
concentrations between $x = 0.8$ and $x = 0.9$.
The ferromagnetic insulating phase (FMI) in the region $x<0.5$
transforms for $x>0.5$ to an AFMI phase.
% with respect to referee A2
A charge ordering (CO) state is found in the concentration range
$0.5 \leq x \leq 0.7$ with a maximal CO transition temperature
of about $\SI{270}{\kelvin}$ at $x=0.5$.
Unlike many doped manganites, there is no 
indication of a metal-insulator transition below the experimental
limit of \SI{9}{\tesla}.

Nevertheless, the underlying microscopic
mechanisms are not yet fully understood: The entire character of 
the magnetic phases is unknown, because Gd and related compounds cannot
be easily investigated by means of Neutron diffraction.
In fact, Gd has shown to be the strongest neutron-absorbent among
 all natural elements \cite{0953-8984-21-12-124201}.

At this point, our theoretical study allows to identify the 
magnetic ground state from total energy calculations
for various potential magnetic phases (see \Fref{fig:magnetic ground states}).

We reexamine at first the two undoped systems GdMnO$_3$ (GMO) and
CaMnO$_3$ (CMO) as a benchmark for our density functional (DFT)
calculations. But when we consider the different concentrations
of the solid solution GCMO, the disorder complicates
the supercell calculations, necessary to cover all magnetic structures given in
 \Fref{fig:magnetic ground states}.

  % with respect to referee B1
On the one hand, disorder could be taken into account
by an effective medium theory -- namely the coherent potential
approximation (CPA) in the framework of  the 
Korringa-Kohn-Rostocker Green's function (KKR-GF) method 
\cite{Hoffmann2012own, Note1}.
Another elegant way to model disordered systems
pioneered by Zunger \etal \cite{PhysRevLett.65.353} 
is the concept 
of special quasi random structures (SQS) for the 
rare-earth site mixed with Ca. We decided to use the latter 
approach because it allows lattice relaxations
and could also cover to some degree short-range order effects, which should be
compared with experimental results later.

The magnetic properties are discussed in terms of magnetic exchange interactions
between the Mn sites. They are then used in a classical Heisenberg model in order
to determine the critical magnetic transition temperatures, which agree
very well with the experimental results \cite{BEIRANVAND2017126}.
As the main result, we obtain the type of magnetic ground states,
which could not be accessed directly from the
magnetoresistance experiments in \cite{BEIRANVAND2017126}. 

% ..........................................................
% SECTION ..................................................
% ..........................................................
\section{Undoped Manganites}%: GMO and CMO} % ................
\label{sec:pureG_C_MO}
A lot of work has already been carried out on the theoretical description
of both endpoint compounds in the GCMO series. We refer the reader
for more details to 
%exist already a large number of
%theoretical works about GdMnO$_{3}$
\cite{PhysRevB.93.075139,Ferreira2018,doi:10.1002/adfm.201604513,
	MEKAM2012156,RAHNAMAYEALIABAD2017942,2018arXiv180502172F} for 
GdMnO$_{3}$ and  
\cite{PhysRevLett.102.117602,doi:10.1063/1.4958716,
PhysRevB.88.054111,PhysRevB.95.115120,
PhysRevB.95.220401,PhysRevB.97.024108} for CaMnO$_{3}$. We aim 
at the beginning to validate the structural, electronic and
magnetic properties  against the previous theoretical and experimental results
as a benchmark for the following discussion of the phase diagram
in \sref{sec:3:phase diagram}.

Our density functional theory (DFT) calculations 
were carried out with the
% with respect to referee A3
projector augmented-wave method 
  \cite{PhysRevB.50.17953,PhysRevB.59.1758}
as implemented
in the Vienna \textit{ab initio} simulation package
(VASP) \cite{KRESSE199615,PhysRevB.54.11169}.
For the treatment of the exchange correlation potential,
we compared four common functionals: 
Perdew-Burke-Ernzerhof (PBE) \cite{PhysRevLett.77.3865},
its revised version for 
solids (PBEsol) \cite{PhysRevLett.77.3865}, 
Perdew-Wang (PW91) \cite{PhysRevB.45.13244},
and Perdew-Zunger (PZ) \cite{PhysRevB.23.5048}. 
An isotropic screened on-site Coulomb interaction 
\cite{PhysRevB.57.1505}
-- the Hubbard $U$ correction --
was added to all aforementioned functionals. 
The choice for the
exchange correlation functional and $U$ was made based on 
the best compromise between the three most important properties:
the electronic band gap, the magnetic moment, and primarily 
the stability of the magnetic order. From those properties,
we considered PBE$+U$ with an $U$ applied on
the Mn $3d$ orbitals with $U_{\text{Mn}}=\SI{2}{\eV}$ as the best choice,
% with respect to referee C2
while the Gd \textit{f}-electrons are treated as
frozen in the core region (motivated by the magnetic
properties, see \sref{sec:heisenberg}).
A thorough discussion and comparison is given in 
the supplemental material \cite{Note2}.

% ..........................................................
\begin{figure*}
	\begin{center}
    \includegraphics[width=480pt] {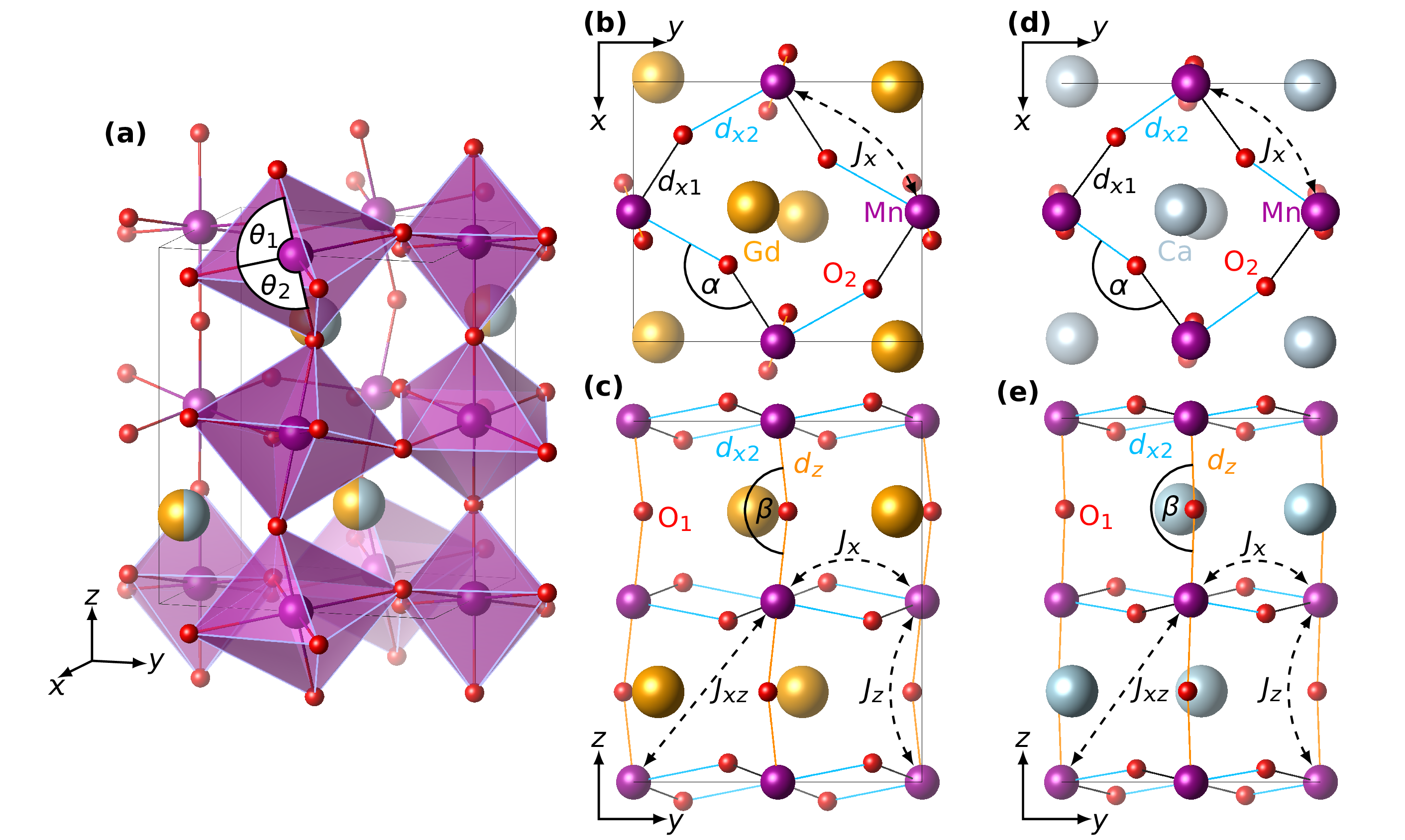}
  \end{center}
	\caption{Structural representation of the Pbnm unit cell of GCMO.
    (a) Schematic of the 
    3-dimensional unit cell including the distorted oxygen octahedra. 
    (b),(c) GdMnO$_{3}$ and (d),(e) CaMnO$_{3}$.
    The colored balls 
    depict the $R$ site (mixed colors), Gd (golden), Ca (gray),
    Mn (violet), and oxygen (red).
    (b),(d) show the
    respective top view ($xy$ plane). (c),(e) feature the side view 
    ($yz$ plane). The structural notation is also indicated for
    bond length Mn--O and the bond angle enclosed in the Mn--O--Mn bond
    ($\alpha$ and $\beta$).
    The three different bond length are noted as \MnOI (orange, $\dOI$),
    \MnOIIa (black, $\dOIIa$), and \MnOIIb (blue, $\dOIIb$).
    The direction of the magnetic exchange interactions between the Mn 
    sites is pictured as well with dashed arrows. 
    The $\theta_{i}$ in (a) represent two of the 
      % with repect to B4
    intra-octahedron bond angles.
    Structural figures were prepared with VESTA \cite{Momma2011jac}.
		}
	\label{fig:structure}
\end{figure*}
% ..........................................................

\subsection{Lattice and electronic structure}
\label{ssec:structure GMO CMO}
Both compounds crystallize in the orthorhombic structure with
the Pbnm symmetry of the space group 62 including 20 sites (\Fref{fig:structure}) 
\cite{MORI2002238,Jorge2005_CMO_expDATA}.
The Gd or Ca atoms occupy the
$4c$ Wyckoff position ($x_R$,$y_R$,$\nicefrac{1}{4}$), 
while the Mn atoms are 
at the $4b$ Wyckoff position ($\nicefrac{1}{2}$,0,0).
The oxygen atoms are located at two different sites and 
are denoted as O$_1$ for $4c$
($x_\text{O$_1$}$,$y_\text{O$_1$}$,$\nicefrac{1}{4}$ )
and O$_2$ for $8d$ ($x_\text{O$_2$}$,$y_\text{O$_2$}$,$z_\text{O$_2$}$)
(see \Tref{tab:pure_str}). The first type of oxygen ions (O$_1$)
forms bonds with the Mn in $z$ direction, while the second type
(O$_2$) is bonded to Mn ions in the ($xy$) plane (see \Fref{fig:structure}).

% TABLE ....................................................
% with respect to referee B2 and B3
\begin{table}
  \caption{Experimental and calculated structural properties
    of GMO and CMO. The lattice constants ($a$, $b$, and $c$) and 
    the bond lengths $\dOI$, $\dOIIa$ and $\dOIIb$
    are given in $\si{\angstrom}$. The latter correspond to \MnOI, 
    \MnOIIa and \MnOIIb, respectively
    (see \Fref{fig:structure}). 
    The Wyckoff positions ($x,y,z$) are given in units of the lattice
    vectors (see text).
    The Baur’s
    distortion index $B_\text{D}$ is dimensionless.
    The angle variance $\sigma^2$ is in (degree)$^2$.
    Note that the $V_0$ calculation scheme
    uses experimental lattice constants \cite{Note3} and the ferromagnetic 
    Mn spin ordering (\sref{sec:heisenberg}).
  }
  \begin{tabular*}
 {\columnwidth}
 {l @{\extracolsep{\fill}} c cc | c cc }
 \hline\hline
 & \multicolumn{3}{c|}{GdMnO$_{3}$} & \multicolumn{3}{c}{CaMnO$_{3}$}\\
 \hline
    & \multicolumn{1}{c}{Exp} & \multicolumn{2}{c|}{DFT} &   \multicolumn{1}{c}{Exp} & \multicolumn{2}{c}{DFT} \\
 \hline
  & \cite{MORI2002238} &  $V_0$ & $V_\text{rlx}$ & 
\add{\cite{Jorge2005_CMO_expDATA}} & $V_0$ & $V_\text{rlx}$ \\
    \hline
$a$              & 5.318 & 5.309 & 5.344 & 5.270 & 5.269 & 5.294 \\
$b$              & 5.866 & 5.852 & 5.937 & 5.279 & 5.284 & 5.332  \\
$c$              & 7.431 & 7.425 & 7.426 &   7.456 &7.457 & 7.496 \\
\\                                                            
$x_{R}$          & 0.938 & 0.981 & 0.981 &   0.990 & 0.992 & 0.992 \\
$y_{R}$          & 0.080 & 0.082 & 0.085 &   0.032 & 0.040 & 0.040  \\
\\                                                      
$x_{\text{O}_1}$ & 0.103 & 0.109 & 0.110 &   0.068 & \add{0.071} & 0.071 \\
$y_{\text{O}_1}$ & 0.471 & 0.465 & 0.465 &   0.493 & \add{0.488} & 0.487 \\
\\                                                      
$x_{\text{O}_2}$ & 0.205 & 0.203 & 0.204 &  \add{0.211} & 0.209 & 0.209 \\
$y_{\text{O}_2}$ & 0.175 & 0.175 & 0.172 &   \add{0.209} & \add{0.210} & 0.210 \\
$z_{\text{O}_2}$ & 0.550 & 0.552 & 0.552 &  0.530 & 0.538 & 0.536 \\
\\                                                      
$\dOI$           & 1.944 & 1.958 & 1.958 & \add{1.891} & \add{1.902} & 1.912 \\
$\dOIIa$         & 1.910 & 1.920 & 1.923 &  \add{1.896} & \add{1.911} & \add{1.923} \\
$\dOIIb$         & 2.228 & 2.224 & 2.265 &   \add{1.907} & \add{1.906} & \add{1.920} \\
\\
$B_\text{D}$     & 0.065 & 0.062 & 0.070 &  \add{0.003} & 0.001 & 0.002 \\
\\                                                      
$\sigma^{2}$     & 3.883 & 5.915 & 6.776 &  \add{0.281} &\add{0.377} & 0.213 \\
    \hline
  \end{tabular*}
  \label{tab:pure_str}
\end{table}
% ..........................................................
  % with respect to referee B2, B3, B4, B5, C1
The orthorhombic structure remains also the lattice structure for
the whole Gd$_{1-x}$Ca$_{x}$MnO$_3$ series for all Ca
concentrations $x$ as the experimental measurements by Beiranvand \etal 
\cite{BEIRANVAND2017126} confirm. We adapt therefore the Pbnm
symmetry in all following calculations, either as the 
primitive cell with 20 sites or as a supercell repeating the 
Pbnm cell $2\times2\times2$ times -- in total 160 sites. 
The latter has to be adapted as stated already above in the introduction 
because we have to take into account all potential 
magnetic spin orientations (\Fref[]{fig:magnetic ground states}),
as well as the disordered character of a solid solution
(see \sref{sec:3:phase diagram}).
As a consequence, the numerical calculation of the volume relaxation 
and the relaxation of the internal coordinates
is too time consuming, because of the large number of sites in the supercell.
Hence, we fix the lattice constants of the Pbmn cell to the 
measured values \cite{BEIRANVAND2017126,Note3}, but 
the internal coordinates could not be accessed by the 
latter references, and had to be obtained by numerical relaxations.
In particular for $x\neq0$ or 1 in GCMO, the experimental internal parameters 
are not yet available. For that reason, we validate our numerical results 
for GMO and CMO against the experimental data in 
\cite{MORI2002238,Jorge2005_CMO_expDATA} (\Tref{tab:pure_str}).
The lattice parameters \cite{BEIRANVAND2017126,Note3} 
are used and the internal coordinates are allowed to relax
\cite{Note2}. This is referred as $V_0$ calculation scheme in
\Tref{tab:pure_str}.

For GMO and CMO,
the internal coordinates vary only slightly from the 
experimentally obtained atomic positions (\Tref{tab:pure_str}).
The resulting Mn--O bond lengths are in good agreement with those
in \cite{MORI2002238,Jorge2005_CMO_expDATA}.
For comparison, we calculated also the full volume relaxation.
The results are marked as $V_\text{rlx}$ calculation scheme in
\Tref{tab:pure_str}.
The GMO volume is slightly overestimated by about \SI{1.6}{\percent} compared 
to experimental values (\Tref{tab:pure_str}). Consecutively, the octahedron
volume was found to be \SI{3}{\percent} larger than in
\cite{MORI2002238}.
Also the overall volume of the CMO cell and its octahedron volume were
found to be overestimated by \SI{1.8}{\percent} and \SI{2}{\percent}, 
respectively. Such overestimation is known as a
characteristic of the GGA functionals in general. The internal parameters
for GMO and CMO agree on the contrary very well between the $V_0$ and 
$V_\text{rlx}$ calculation schemes (\Tref{tab:pure_str}). This motivated again 
the choice of the experimental lattice constants \cite{Note3}.

Furthermore, both experimental references 
\cite{MORI2002238,Jorge2005_CMO_expDATA}
show the characteristic manganite lattice 
distortions as described in the introduction. 
The deviations from the ideal cubic
perovskite can be quantified using the two angles, 
$\alpha$ and $\beta$ (\Fref{fig:structure}), 
the Baur’s
distortion index ($B_\text{D}$) \cite{Baur:a11025},
and the bond angle
variance ($\sigma^{2}$) \cite{Robinson567}. 
 $B_\text{D}$ expresses the deviations of the
Mn--O distances from their mean value. 
In an undistorted octahedron, the
three Mn--O distances are equivalent and $B_\text{D}$ is zero.
% with respect to referee B4
The bond angle variance measures the deviation of the twelve O--Mn--O
intra-octahedron bond angles $\theta_i$ (\Fref{fig:structure}a)
from those \SI{90}{\degree} bond angles in an ideal 
octahedron of the same volume. Thus, $\sigma^2$ becomes zero for the ideal 
octahedron. In contrast  to the $\theta_i$ the angles $\alpha$ and $\beta$  
quantify the mutual tilting of the octahedra.

A similar agreement between experimental and theoretical results,  
observed for the internal parameters above, is reflected in the
distortion index and the bond angle variance as well. We observe an almost 
similar 
$B_\text{D}$ for GMO but some deviations in $\sigma^2$ for GMO and CMO.
This shows the advantage of considering both quantities, because they 
highlight differences in the internal coordinates otherwise not obvious.
Nonetheless, we obtain a reduction of the octahedron distortion
expected from experimental studies, which is visible in the 
three Mn--O bond length -- almost similar in CMO, but different in GMO.
% with respect to referee B5
We conclude that the structural distortions in GMO and CMO
\cite{MORI2002238,Jorge2005_CMO_expDATA}
 are well resembled by the 
atomic coordinates and the distortion indices determined in
our DFT calculations (\Tref{tab:pure_str}). 

Considering the electronic structure, 
we obtained an insulating state for both compounds -- GMO and CMO 
\cite{Note2}.
Our calculated Kohn-Sham band gap of GMO without correlation corrections (\SI{0.38}{\eV})
agreed with the result by 
Kov\'a\ifmmode \check{c}\else \v{c}\fi{}ik \etal \cite{PhysRevB.93.075139}. 
It increases to \SI{1.1}{\electronvolt} with our choice 
of $U = \SI{2}{\electronvolt}$ \cite{Note2},
although it is still below experimental band gaps
obtained from UV absorption spectra of GMO nanoparticles (\SI{2.0}{\electronvolt}) 
\cite{doi:10.1063/1.3358007} or optical measurements
(\SI{2.9}{\electronvolt}) \cite{Negi2013}.  We observe
a strong hybridization between $\eg$ states and O $p$ 
states at the valence band maximum
of the A-AFM ground state of GMO (see discussion of magnetic
ground states below), while the conduction band minimum is 
formed by a notable mixing between Mn 
  % with respect to referee A6
  $\eg$ and $\tg$ states \cite{Note2}.
The $\eg$-like valence band width is \SI{0.95}{\electronvolt}, in 
line with the reported GW band structure \cite{PhysRevB.93.075139}. 

We obtain similar features for the calculated band gap of CMO obtained 
with PBE$+U$, which is \SI{0.92}{\electronvolt}. This value is again
lower than the experimental band gap (\SI{1.55}{\electronvolt})
measured for single crystals of CMO \cite{PhysRevB.70.224406}.

% ..........................................................
\subsection{Mapping to a classical Heisenberg model and determination of transition temperatures}
\label{sec:heisenberg}

% GENERAL PROPERTIES of Gd and  rare- earth manganites
% with resepct to referee C2
Of all rare-earth elements,
Gd stands out due to unique properties. 
The Gd$^{3+}$ ions 
have the largest magnetic moment 
caused by 7 unpaired spins and show in GdMnO$_3$
the largest observed ordering temperature (\SI{6.5}{\kelvin}
\cite{PhysRevB.70.024414}) of the $R$ sublattice in all $R$MnO$_3$ compounds.
Nevertheless, the latter ordering temperature following from 
Gd-Gd magnetic interactions is lower than the one of
the Mn sublattice (\SI{45}{\kelvin} \cite{BEIRANVAND2017126}),
while we can also assume that the transition temperature of the Gd sublattice
will not increase with increasing the Ca concentration. The mean distance 
between the Gd ions will only increase leading to an even weaker magnetic
coupling.
Hence, the Gd-Gd interactions are negligible against the 
magnetic interaction between the Mn ions.
We restrict ourselves to the magnetic ordering of the
the Mn ions only, while the \textit{f}-states  are in the core region. 
Consequently, when we speak 
in the following about a magnetic order, we only refer to the
orientation of the Mn magnetic moments. 

% SUPERCELL 160 atoms 32 f.u. 2x2x2
% with respect to referee A11
In order to identify the magnetic ground state 
structures, we took into account ferromagnetic (FM),
ferrimagnetic (FiM), and antiferromagnetic structures
(A-type, G-type, ... AFM). They are illustrated in
\Fref{fig:magnetic ground states} in their minimal 
required cell, but we needed for the actual calculation
a common supercell to accommodate all possible magnetic configurations.
We used therefore the 160 atoms supercell described above 
with fixed lattice constants \cite{BEIRANVAND2017126, Note3}.
The internal coordinates were relaxed only for 
the FM spin configuration, while they had to be static 
for other magnetic configurations because of the
calculations of the magnetic exchange interactions below.
This assumption may slightly bias our results towards a FM ordering
(see below)
but is a compromise between using the Heisenberg model, much longer
computation time, and
too many other potential sources of changed materials properties --
besides lattice constants, different spin orientations, or later Ca doping.

The total energies are then calculated within this fixed structure 
for different Mn spin orientations. Those magnetic structures
with the lowest total energy resemble the magnetic ground state.
The static FM internal coordinates might bias our results slightly 
towards an FM 
configuration as the ground state but also the 
experimental study observed an FM signal over
a large Ca concentration range \cite{BEIRANVAND2017126},
while several other antiferromagnetic ground states could not be finally
excluded.
%We stay with the fixed structure due to the large cell size and
%primarily because we did not find any qualitative changes
%in the total energy differences for the respective magnetic 
%% with respect to referee A4
%structures when using the structure obtained
%  by
%relaxing with 
%the A-AFM spin configuration.

The relative total energies can then be used to verify
the experimentally found ground states
\cite{PhysRevB.68.060403, Jorge2005_CMO_expDATA}. 
We identified the A-AFM and G-AFM 
as those magnetic structures with the lowest 
total energy for GMO and CMO, respectively
\cite{Note2}.
However, those ground states are not very stable
against magnetic variations, since in both cases
other magnetic structures are close in energy
(FM for GMO, C-AFM for CMO) (see also \Fref{fig:mag phase diagram GCMO}a).

% ..........................................................
\begin{figure}
  \includegraphics[width=246pt] {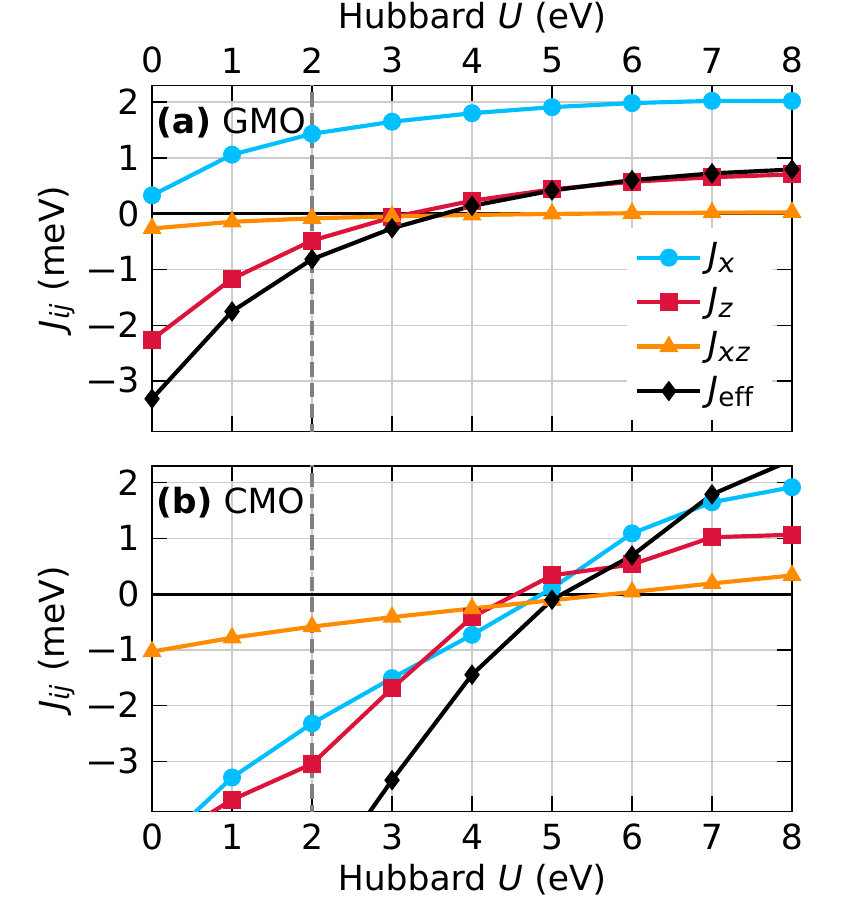}
  \caption{The three Heisenberg exchange interactions
  		% with resepct to C3
  		and $J_\text{eff}=J_z+4 J_{xz}$
  	in dependence 
    of the correlation treatment in (a) GdMnO$_{3}$ and (b) CaMnO$_3$. 
    The gray dashed line indicates the choice of $U=\SI{2}{\eV}$ in 
    this work. 
    See \Fref{fig:structure} for the visualization of 
    the three magnetic coupling directions.
}
  \label{fig:JsRMnO3}
\end{figure}
% ..........................................................

% Magnetic moments
As mentioned above in the introduction, the Mn ion appears 
in two different valence states for GMO (Mn$^{3+}$) and CMO (Mn$^{4+}$)
due to the different valence electron configuration of Gd and Ca.
We obtain from our DFT calculations magnetic moments of
$3.6\mu_\text{B}$ for Mn$^{3+}$ (GMO) and
$2.7\mu_\text{B}$ for Mn$^{4+}$ (CMO) 
% with respect to referee A5
  \cite{Note2}, which substantially deviate from their 
integral value of $4\mu_\text{B}$ and $3\mu_\text{B}$, 
respectively. In GMO, this deviation is 
caused by the aforementioned hybridization of the 
Mn states with the oxygen states introducing also 
a magnetic moment of 0.06$\mu_\text{B}$ 
  % with respect to referee A6
  at the oxygen ions \cite{Note2}.
Our observed local magnetic moment of CMO is 
in line with its experimental value of
$2.665\mu_\text{B}$ \cite{PhysRev.100.545}, while we
did not find any experimental value of the
local magnetic moment of Mn$^{3+}$ in GMO. Nevertheless,
the $3.6\mu_\text{B}$ for Mn$^{3+}$ in GMO
agree with earlier numerical calculations 
including hybrid functionals \cite{PhysRevB.93.075139}. 

% JIJ DISCUSSION
In addition to the magnetic ground state, we will need below
for a full description of the magnetic phase diagram of GCMO
also the corresponding finite temperature characteristics
-- namely the critical transition temperatures. 
The latter can be derived on basis of the classical 
Heisenberg model from DFT total energies.
% with resepct to referee B6
Therefore, the total energies are mapped onto a Hamiltonian of the form
\begin{equation}
	H = - \frac{1}{2} \sum_{i \neq j} J_{ij} \,  \vec{S}_i \vdot \vec{S}_j\,.
	\label{eq:Heisenberg}
\end{equation}
The parameters $J_{ij}$ are $J_x$ (in-plane interaction), $J_z$ 
(out-of-plane interaction), and $J_{xz}$ (interaction along the cell diagonal), 
if $(ij)$ describes a corresponding pair of atoms as indicated in
\Fref{fig:structure}.
% with resepct to B13
We can also define an effective 
out-of-plane interaction $J_\text{eff}=J_z+4 J_{xz}$, which can characterize 
the tendency to an out-of-plane antiferromagnetic order \cite{Note2}.
The sums in 
\eqref{eq:Heisenberg} run over all sites $i$ with the 
interaction sites corresponding to each $J_{ij}$.
Positive (negative) $J_{ij}$ correspond to FM (AFM) coupling.
The spin moment $S$ in \eqref{eq:Heisenberg} 
equals to 2 for Mn$^{3+}$ (4 unpaired electrons/2) 
and $\nicefrac{3}{2}$ for Mn$^{4+}$ (3 unpaired electrons/2).
This kind of Hamiltonian was used to study magnetic 
properties of GMO before \cite{PhysRevB.93.075139} and
has an advantage over many other studies on magnetic properties 
of $R$MnO$_{3}$ being restricted only to the nearest 
Mn neighbors exchange couplings.

The three magnetic exchange parameters can be then 
obtained by mapping total energies of different spin orientations
(\Fref{fig:magnetic ground states}) onto the Heisenberg 
Hamiltonian in \eqref{eq:Heisenberg}. This results in an 
over-determined set of equations, which is solved with 
	%with respect to referee B7
	% cite also the Supp Mat
	a linear least square fit \cite{Note2}.
The ferrimagnetic configuration FiM 
(\Fref{fig:magnetic ground states}) was used as the 
reference energy $E_{0}$ inspired by \cite{PhysRevB.93.075139}.

At this point, we want to emphasize again the importance
of a correct electronic correlation treatment in our materials.
Our determined magnetic exchange interactions vary strongly
with increasing $U$ parameters (\Fref{fig:JsRMnO3}).
We even obtained with the PBE exchange correlation functional, without $U$ correction,
	for GMO a wrong G-type AFM ground state.
% with respect to B13 & C3
The out-of-plane contribution characterized by $J_\text{eff}$
dominates the in-plane interaction (\Fref{fig:JsRMnO3}a).
Only when $U$ is increased to be around \SI{2}{\eV}, 
the in-plane exchange becomes stronger and  
leads with $J_x>0$ and $J_\text{eff}<0$ to the correct 
known A-AFM phase \cite{Note2}.
Increasing $U$ further results in an FM order: 
	% with respect to C3
the magnitude of $J_\text{eff}$ decreases and it turns 
positive (ferromagnetic) for $U\gtrsim\SI{4}{\eV}$  (\Fref{fig:JsRMnO3}a).
This observation matches well with the potential instability
against a FM state found for GMO based on
the total energy calculations. The energy 
difference between the FM and A-AFM states is $\sim\SI{4}{\meV}$ 
(\Fref{fig:mag phase diagram GCMO}).

For CMO, the situation is slightly different.
The three exchange interactions are negative for plain PBE 
(see \Fref{fig:JsRMnO3}b) and only become positive
for $U>\SI{5}{\eV}$, which is far above a reasonable value
considering other materials properties. The strong competition
between the three exchange parameters for $U<\SI{4}{\eV}$
leads to the G-type AFM phase.

Finally, we want to assess the magnetic transition temperature
(either $T_{\text{N}}$ for AFM phases or $T_{\text{C}}$ for FM and FiM)
and used our own Monte-Carlo 
(MC) code \cite{PhysRevB.80.014408}
together with the Heisenberg Hamiltonian \eqref{eq:Heisenberg}.
Therein, we use a large cluster of $16\times 16 \times 16$ times
the primitive unit cell (a total volume of about \SI{100}{\angstrom\cubed}).
Periodic boundary conditions are also considered.
The thermal equilibrium was firstly assumed to be reached after 
\num{60000} MC steps. Another \num{20000} steps are then used
in the thermal averaging. We started from a high temperature
of $\SI{500}{\kelvin}$  and cooled down the GCMO samples in steps of 
 $\SI{3}{\kelvin}$. The transition temperatures are later extracted
 from the temperature dependence of three quantities -- the
 magnetic susceptibility, saturation magnetization, and the heat capacity.   
The calculated exchange interactions are used for the
initial system configuration.
An ordering temperature of \SI{42}{\kelvin} was obtained for GMO,
which matches perfectly the experimental value of \SI{40}{\kelvin} 
\cite{PhysRevB.68.060403,PhysRevB.70.024414}. In contrast,
a hybrid functional calculation led to a little overestimation of 
$T_{\text{N}}$ by about $\SI{20}{\kelvin}$ \cite{PhysRevB.93.075139}.
For CMO, a $T_{\text{N}}$ of $\SI{96}{\kelvin} $ was obtained,
which is in the same range as the experimentally
observed $T_{\text{N}}$ of $\SI{125}{\kelvin} $ \cite{PhysRevB.59.8784}. 

We conclude that our computational setup and the 
procedures in order to obtain the magnetic ground state and the 
magnetic transition temperatures produce results in good agreement
with available experimental data. Therefore, we have
a proper basis for the study of the complete series of 
intermixed rare-earth and alkaline earth manganites.
% ..........................................................
% SECTION ..................................................
% ..........................................................
\section{Phase diagram for GCMO}
\label{sec:3:phase diagram}

% Structural features together with experiments.
Using the orthorhombic structure for all Ca concentrations $x$ in the 
Gd${_{1-x}}$Ca${_{x}}$MnO$_{3}$ series,
we want to take into account the disordered character of
this solid solution.
Besides the above mentioned KKR-CPA method, another
possible way is to average the (Gd,Ca) sublattice occupancy
over different structures within a large supercell with $N$ functional units.
Such method is impractical because one has in general to average
	% with resepct to referee B8
about too many
configurations, even if symmetry arguments are used to find potentially 
different configurations.
In order to circumvent the problem, 
we used the special quasi-random structure (SQS) 
method \cite{PhysRevLett.65.353} for the (Gd,Ca) sublattice.
SQS takes into account the random nature of alloys by choosing the 
occupation of the internal coordinates inside a supercell in such a 
way that the pair and
multi-site correlations mimic as much as possible those 
of a random substitutional alloy.
The multi-site correlations in the SQS candidates 
are then taken into account and compared to the random distribution
up to a defined cutoff radius.   
% with respect to referee A8
The resulting structure based on these constraints is not
necessarily a fully disordered structure, but a good 
approximation to the real solid solution character of the
material, as follows from the correlation functions of the SQS
\cite{Note2}.

To the best of our knowledge, it is the first time that
the SQS method is applied to such manganites.
The generation of the SQS cells was
carried out using a Monte-Carlo annealing loop, 
as implemented in the MCSQS routine of the ATAT package \cite{VANDEWALLE201313}.
We forced the axis orthogonality in the SQS cells,
which kept the distance between the Mn sites and mostly the
angles between them constrained throughout the GCMO series.
In this way, we can keep the same definition of the three 
aforementioned exchange coupling constants defined in \sref{sec:heisenberg}
for the following comparison of magnetic properties 
throughout the concentration range.

Nevertheless, the concentration $x$ cannot be chosen
continuously between \SIrange{0}{100}{\percent} but 
depends on the size of the 160 atoms supercell. 
Therefore, the smallest concentration step
used in the simulation can be only $1/2^{3}=\nicefrac{1}{8}$
and we performed all calculations for the concentrations
$x=0$, $\nicefrac{1}{8}$, $\nicefrac{1}{4}$, ..., 
$\nicefrac{7}{8}$, 1.% , and 1 ($i=1,...,9$).

Many SQS reported in the literature are obtained by matching
only pair correlations.
In this work, we include also higher order correlations of the
random structure. Pair clusters are taken up to the \nth{5}
nearest neighbor, triplet and quadruplet clusters are included up to
the \nth{4} nearest neighbor. Only for $x=0.5$,
the SQS structure fully resembles a completely disordered system
with zero correlation functions (see \cite{Note2}). The other correlation
function results and structural details of the SQS 
are collected in \cite{Note2}.
%%%% DOS

% with respect to referee A10/A11
Although we include also disordered Ca doping in our study,
there are still some limitations.
A careful consideration of the
coupling between spin, charge, lattice and orbital degress of freedom
should be done, but this is far from trivial, especially
with our large supercells. On top of the additional 
plethora of relaxations for the AFM structures, different
charge or orbital ordering states should be taken into account
-- maybe even in dependence of different Ca/Gd distributions.
Such a number of correlations is beyond the scope of this work and
we restrict therefore ourselves to the coupling between
the variation of lattice constants, the disorder-like Ca concentration within 
an SQS cell, and the magnetic states considered for static internal parameters.

% EXPERIMENTAL LATTICE CONSTANTS:
% could be maybe used in connection with the bond angles etc. 
The experimental lattice parameter $a$ and $c/\sqrt{2}$ vary only 
little with a slight maximum for $x=0.5$. Only $b$ decreases 
strongly until $x=0.5$ and follows afterwards $a$ and $c/\sqrt{2}$
\cite{BEIRANVAND2017126, Note3}.
That means that the unit cell volume of GCMO contracts 
with increasing Ca concentration $x$, which results 
in a gain of Mn$^{4+}$ content.
Such volume contraction in manganites is commonly explained
based on ionic radii, 
because the ionic radius of 
6-fold coordinated Mn$^{4+}$ (\SI{0.53}{\angstrom}) is smaller compared to that 
of Mn$^{3+}$ (\SI{0.645}{\angstrom}) \cite{Note4}.
But at the same time, the ionic radius of the introduced Ca$^{2+}$ ions
(12-fold coordinated: \SI{1.34}{\angstrom}) is larger than that of 
the substituted 
Gd$^{3+}$ ions' (9-fold coordinated: $\SI{1.107}{\angstrom}$).
The 9-fold coordination is a good estimate due to 
the strong distortions in GMO (see Fig. 2).
For Ca$^{2+}$, the coordination is as well not clearly 12-fold
but also the ionic radius of 
10-fold coordinated Ca$^{2+}$ (\SI{1.23}{\angstrom})
is still larger than the one of Gd$^{3+}$ and the 
above statement holds true. Despite
all these aspects, an overall volume contraction is still observed 
together with less distortions in the Mn octahedra.
Thus, a simple analysis based on ionic radii alone is not 
possible but we see that several structural aspects are intertwined:
ionic radii, site coordinate, doping concentration, and atomic bonding.

% ..........................................................
\begin{figure}
	\centering
	\includegraphics[width=246pt]{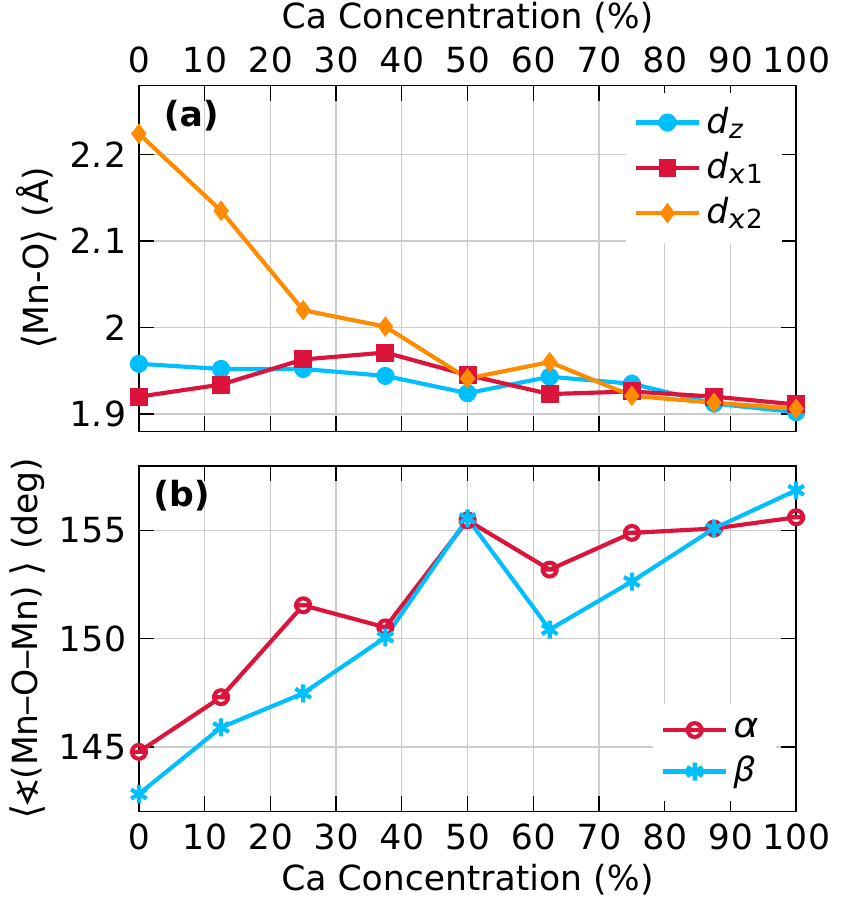}
	\caption{The variation of (a) the three Mn--O bond lengths
		and (b) the Mn--O--Mn bond angles averaged
	 over the SQS bond lengths and angles, respectively, with respect to the Ca 
    concentration in the whole GCMO series. 
    See \Fref{fig:structure} for the visualization of both 
    structural properties.
  }
	\label{fig:bondsgcmo}
\end{figure}
% ..........................................................

We tracked the distortion
of the Mn octahedra via the variation of bond angles and bond lengths
in all GCMO compounds. Therefore,
we calculated the mean value of all present bond lengths (angles)
inside the relaxed SQS cells (\Fref{fig:bondsgcmo}).
The changes of the bond lengths match the 
behavior of the experimental lattice constants
in having distinct changes at $x=0.5$ (\Fref{fig:bondsgcmo}a),
which holds also true for the Mn--O--Mn bond angles (\Fref{fig:bondsgcmo}b).
The resulting distortion indices, $B_\text{D}$ and $\sigma^{2}$
(not shown), decrease linearly 
with increasing $x$. Only
at $x = \nicefrac{5}{8}$, both indices show an anomaly,
which follows exactly the peculiar deviation of the cell
parameters at the aforementioned concentration.
  
The calculated density of states of the GCMO series shows
  % with respect to referee A12, 
  % and adjusted for B9
a half-metallic-like behavior, that means the DOS being metallic in the 
majority spin channel, but having a band gap
of \SIrange{1}{1.5}{\electronvolt}
in the minority spin channel (see \cite{Note2}).
A similar result was shown for La$_{1-x}$Ca$_{x}$MnO$_{3}$
\cite{PhysRevB.89.205110}, where 
the insulating character of the density of states was only recovered
by localizing the additional electron (hole) in the system. 

The magnetic ground state structures for GCMO are determined,
as in section \ref{sec:heisenberg},
for the SQS at every concentration as well.
The number of
relevant magnetic exchange interactions remains also the same,
$J_x$, $J_z$, and $J_{xz}$ (\Fref{fig:structure}),  
due to the conserved Mn distances in the supercells. 
We only vary $S$ as
the mean value of the spin moments, which 
corresponds to the respective Ca concentration
\begin{align}
  S_{x}= (1-x)S_{\text{Mn}^{3+}} + x S_{\text{Mn}^{4+}}\,,
  \label{eq:spin moment average}
\end{align}
with $S_{\text{Mn}^{3+}}=2$ and 
$S_{\text{Mn}^{4+}}=\nicefrac{3}{2}$.
%The same approach was adapted to the oxidation state of 
%the Mn ions.
In the case of partial occupation 
of Gd sublattice ($0 < x < 1$), 
the distinction  between % the quantum spin numbers of
Mn$^{3+}$ and Mn$^{4+}$ is ignored in all our calculations. 
They are
treated at the same footing as effective Mn ions 
with concentration dependent valence states 
taking a value of $3+$ at $x = 0$ and $4+$ in $x = 1$. 
Following the experimental literature 
\cite{BEIRANVAND2017126,PhysRev.100.545,PhysRev.100.564},
we can distinguish three different doping regimes: hole doping
for $x<\nicefrac{1}{2}$, middle doping region for $\nicefrac{1}{2} \leq x < \nicefrac{7}{8}$, 
and electron doping $x\geq \nicefrac{7}{8}$. 
% with respect to B10
For $x<0.5$ the lattice parameter $b$ is considerably larger than 
$a$, but they become equal for $x\geq 0.5$ \cite{BEIRANVAND2017126}.
Connected to the change in 
the lattice constants, all Mn-O bond lengths become nearly equal beginning from 
$x=0.5$, while the tilting angle $\alpha$ becomes practically independent on
the concentration in this region (\Fref{fig:bondsgcmo}). 

% ..........................................................
\begin{figure}
	\includegraphics[width=246pt]{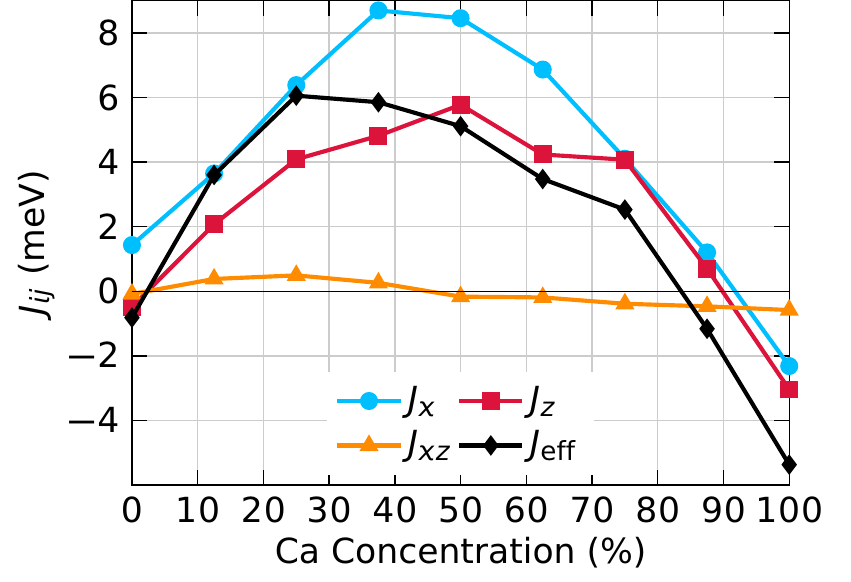}
	\caption{The calculated Heisenberg exchange 
		interactions in Gd${_{1-x}}$Ca${_{x}}$MnO$_{3}$
		following equations \eqref{eq:Heisenberg} and
		\eqref{eq:spin moment average}, 
	% with respect to referee B13
		and %the effective 
%		out-of-plane interaction
		$J_\text{eff}=J_z+4 J_{xz}$.
		See \Fref{fig:structure} for the visualization of 
		the three magnetic coupling directions.
% with respect to referee A7
    $U=\SI{2}{\eV}$ is used in the underlying electronic
    structure calculations
    \cite{Note2}.
	}
	\label{fig:jgcmo}
\end{figure}
%...................................
% ..........................................................
% ..........................................................
% ..........................................................
% ..........................................................
\subsection{Hole doping:  $x < \nicefrac{1}{2}$ }
Adding Ca to GdMnO$_3$ introduces a hole in the vicinity of
the Ca$^{2+}$ ion, which is compensated by an additional electron
from Mn -- the already mentioned Mn$^{4+}$ is created. 
This process causes a transition of the A-AFM phase 
to a FM state in the concentration range $0 < x < 0.5$,
 experimentally verified by
Beiranvand \etal \cite{BEIRANVAND2017126}.
Their temperature dependent SQUID measurements
show in addition a negative magnetization at $x = 0.1$ and
 $T < \SI{20}{\kelvin}$, which they
mainly attributed to the Gd spins -- orienting themselves
antiparallel to the direction of the Mn spins.
This ferrimagnetic coupling was firstly proposed 
for $x = 0.3$ \cite{PhysRevB.55.6453} and thereafter generalized for
% all doping levels 
 $x < \nicefrac{1}{2}$ \cite{B202079N,BEIRANVAND2017126} of GCMO. 
% ..........................................................
\begin{figure}
	\centering
	\includegraphics[width=246pt]{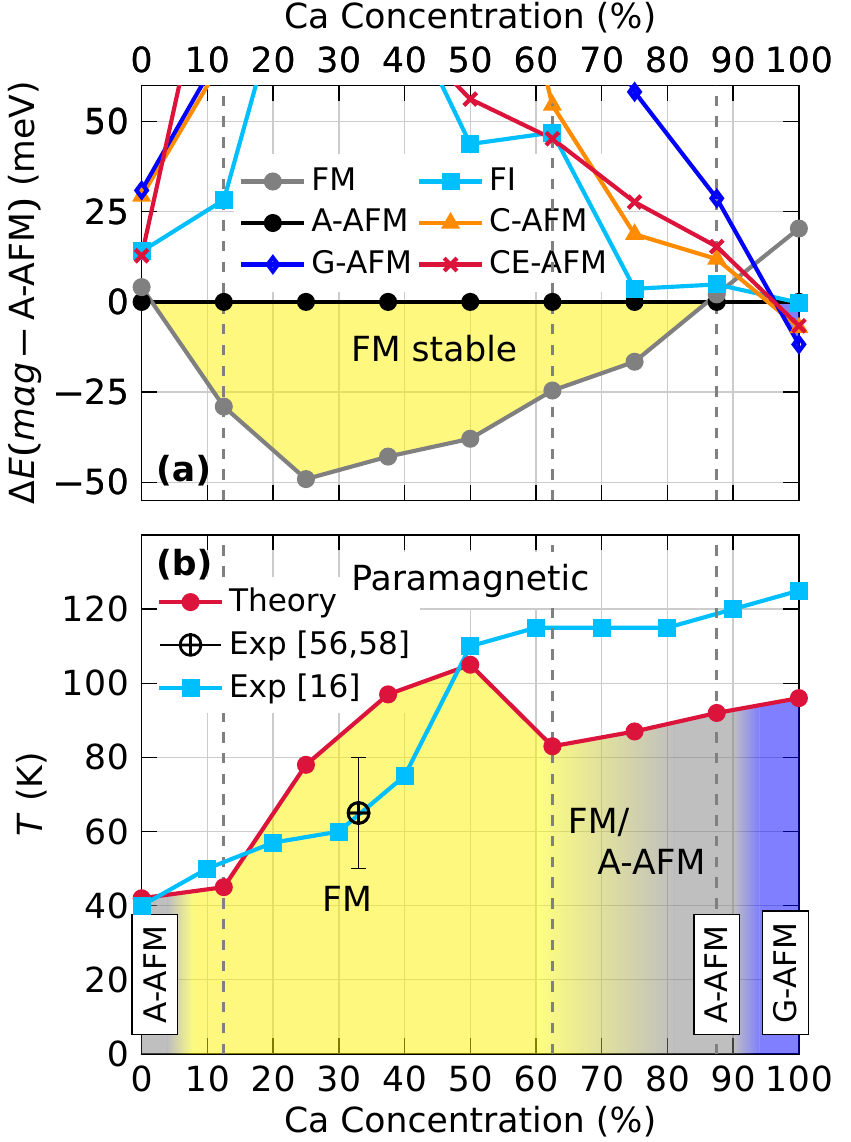}
	\caption{(a) The concentration dependent total energy 
		landscape of the most relevant magnetic ground state structures ($\mathit{mag}$) 
		depicted in \Fref{fig:magnetic ground states}. 
		The energy differences $\Delta E(\mathit{mag} - \text{A-AFM})$ are calculated 
		with respect to the A-AFM state. 
    	% with respect to B12
    That magnetic state ($\mathit{mag}$),
		which has the lowest $\Delta E$, is the most stable one.
		(b) The theoretical magnetic phase diagram 
		of Gd${_{1-x}}$Ca${_{x}}$MnO$_{3}$. The critical temperatures
		(red circles) were determined via the Monte Carlo simulations, while the 
		magnetic phases were identified from the minimal total energy. 
		The measured critical temperatures from Ref. 
		\cite{BEIRANVAND2017126} (blue squares) and 
		for $x=\nicefrac{1}{3}$ from Ref. \cite{PhysRevB.55.6453,PENA2007339}
		(black $\oplus$ with error bars)
		are shown for comparison.
		Above the critical temperature, we expect a paramagnetic state.
    % with respect to referee B12
    The region marked with FM/A-AFM identifies the concentration range,
    where the total energy difference of the FM and A-AFM magnetic phase is 
    below \SI{25}{\meV}. Dashed lines mark qualitative changes of the magnetic 
    ordering. 
  }
	\label{fig:mag phase diagram GCMO}
\end{figure}
% ..........................................................
The same FM phase transition is obtained 
in our calculation with the
SQS structure at $x=\nicefrac{1}{8}$.
Before at $x=0$, the FM state has not the lowest total
energy but its energy difference to the A-AFM state
is rather small
(see \Fref{fig:mag phase diagram GCMO}a).
The increase of the Ca concentration to $x=\nicefrac{1}{8}$ turns
the sign of the total energy difference and enhances it strongly: 
the FM state is (\SI{29}{\meV}) below the A-AFM state and even
more for $x=\nicefrac{1}{4}$ (see \Fref{fig:mag phase diagram GCMO}a).
This first transition is connected with a strong increase of the 
in-plane exchange parameter ($J_x$) to
	% with respect to B11
	\SI{3.6}{\meV}
and 
an AFM to FM change of the out-of-plane 
exchange interactions -- 
% with resepct to B13
visible in $J_\text{eff}$
(\Fref{fig:jgcmo}).
The latter goes from negative to positive. An A-AFM state is only 
realized for $J_x>0$ and $J_\text{eff}<0$.
This variation
in the magnetic coupling strength
does not only result in the A-AFM to FM transition but
also in an increased Curie temperature until $x=\nicefrac{1}{2}$
(see \Fref{fig:mag phase diagram GCMO}b),
which qualitatively matches the experimental measurements 
of a FM order in the whole hole-doped region of GMO
\cite{BEIRANVAND2017126,PENA2007339,PhysRevB.55.6453}. 
Such magnetic alteration could be
attributed to the progressive increase of the Mn--O--Mn bond angle with
the doping level, as well as the drastic shrink of 
the in-plane \MnOIIb bond length (\Fref{fig:bondsgcmo}). 
The Mn--O--Mn bond angle was, e.g., reported for 
$x=\nicefrac{1}{4}$ as $\SI{149.7}{\degree}$ \cite{PEKALA200712}, 
which is the average of our two calculated angles,
$\SI{147}{\degree}$ and $\SI{151.7}{\degree}$. 
Accompanied with the decrease of the cell parameter $b$,
the overall cell distortion diminishes and
we can conclude that the Ca induced
magnetic transformation is mainly triggered by the 
reduction of the Jahn-Teller distortion.
The disagreement between the measured and calculated transition
temperatures in \Fref{fig:mag phase diagram GCMO}b could have,
besides the known problems of $T_\text{C}$ calculations,
several different explanations: Lattice imperfections like
vacancies, in particular at the oxygen sublattice, 
might cause significant changes in the magnetic properties
as observed for other oxides, like SrCoO$_3$
\cite{Hoffmann2015prb_own} or Sr$_2$FeMoO$_6$ 
\cite{Hoffmann2017jpcm_own}. 
% with respect to referee A13
In addition, differences between experiments and theoretical
simulations may be connected to the fact, that an ideal periodic crystal is
assumed in the simulation, while the samples are polycrystalline.
Furthermore, a more complicated magnetic structure might occur for $x=0.33$
(canted antiferromagnetic) instead of a simple ferromagnetic state as supposed
by Snyder \textit{et al.} \cite{PhysRevB.55.6453}.
%
% ..........................................................
\subsection{Half occupied: $\nicefrac{1}{2} \leq x < \nicefrac{7}{8}$}
In the mid-doped region for $x \sim 0.5$, 
our Monte-Carlo simulation determined a transition temperature
of \SI{105}{\kelvin} -- close to the reported bulk temperature (\SI{107}{\kelvin}).
The exchange coupling $J_{xz}$ becomes negative
already for $ x = 0.5$
(see \Fref{fig:jgcmo}), but the FM order still remains the ground state. 
With increasing Ca concentration, the energy difference between the FM and 
A-AFM order becomes gradually smaller
(see \Fref{fig:mag phase diagram GCMO}).

The Mn--O--Mn bond angles become equivalent -- both are \SI{155}{\degree}
(\Fref{fig:bondsgcmo}b) and all Mn--O distances decrease to roughly the similar distance
(\Fref{fig:bondsgcmo}a). Hence, the octahedron distortion 
becomes less pronounced than before, which hints also 
to the ferromagnetic order due to the double exchange mechanism
following from the different valency of the Mn ions.

% with respect to referee A14
The concentration $x=0.5$ marks the transition to an
antiferromagnetic ground state in the experimental phase diagram
\cite{BEIRANVAND2017126}. Due to
missing neutron diffraction data the particular type of antiferromagnetic order
is not known from experiments. The calculation still leads to a ferromagnetic
ground state for this concentration, but charge order (CO) is observed
experimentally.
Thus, in a next step of our investigations the concentration range
$1/2 \leq x \leq 7/8$ has to be investigated with inclusion of charge order
phenomena.
This room temperature CO state makes the mid-doped concentration range 
not only most interesting for technical applications 
but might have also an important role in the stabilization of the  AFM order,
which was discussed, e.g., for La${_{0.5}}$Ca${_{0.5}}$MnO$_{3}$ 
\cite{PhysRevLett.75.3336}.
The latter compound was reported to be a
ferromagnetic metal due to double exchange coupling 
but becomes a antiferromagnetic insulator for temperatures
$T\lesssim \SI{195}{\kelvin}$. 
The authors of Ref. \cite{PhysRevLett.75.3336} suggest that
the latter AFM phase transition  coincides with a charge
ordering transition, which suppresses the ferromagnetism
and stabilizes the AFM order.

Another potential stabilization mechanism of the AFM order was proposed
for Pr${_{1-x}}$Ca${_{x}}$MnO$_{3}$ (PCMO). Its magnetic order
at $ x = 0.5$ is rather maintained by the presence
of the so called Zener polarons,
because a stabilization of a CE-type AFM order by means of the CO
could be excluded based on single-crystal neutron diffraction 
measurements  \cite{PhysRevLett.89.097205,PhysRevB.92.155148}.
This phenomenon results from trapped electrons between the two Mn sites
causing a valence of 3.5+ in the neighboring Mn ions instead of the natural
valence of 3+ or 4+, respectively.

An analogous argument was given by Garc\'ia \etal
\cite{PhysRevLett.85.3720} using a ferromagnetic Kondo lattice model.
Therewith, they demonstrate that the formation of magnetic polarons
is an important ingredient in the description of systems with correlated
spin-charge degrees of freedom. This correlation is induced from 
the strong competition between double exchange and superexchange mechanisms.
% with respect to referee A15
The signature of such coexisting spin and charge ordering could be
as well obtained from DFT calculations. However, such phenomena require much 
more computational effort, i.e. different charge patterns have to be checked at 
each $x$ concentration and for all considered magnetic orders in this study.

Adding then more Ca does not change the qualitative picture.
The FM order remains still the 
lowest magnetic ground state structure and
the corresponding Curie temperatures are still high ($>\SI{80}{\kelvin}$)
(\Fref{fig:mag phase diagram GCMO}). However, the total energy 
difference to the A-AFM order is strongly reduced and at
$x=\nicefrac{3}{4}$, the ferrimagnetic (FiM) order 
(\Fref{fig:magnetic ground states}) starts to compete for
the lowest total energy. Here,
$J_{x}$ and $J_{z}$ are equivalent, 
while the AFM coupling $J_{xz}$ increases
(\Fref{fig:jgcmo}).

%The presence of FM polarons in the AFM ground state
%is shown to be inevitable in the electron-doped manganese
%perovskites \cite{PhysRevB.58.R14689}. 
%The stabilization of the CE-AFM order could be also  
%maintained by the orbital ordering \cite{PhysRev.100.545}.
%unluckily none of the mentioned mechanisms could be confirmed with
%our calculations. Both charge and orbital ordering
%effects are not taken into account in our Heisenberg model. 
%
% ..........................................................
% referee B14
\subsection{Electron doping: $\nicefrac{7}{8} \leq x < 1$ }
\label{sec:magnetic_GCMO}

The last doping regime represents essentially CaMnO$_3$ 
doped with few percent Gd ions, which adds excess electrons from 
Gd$^{3+}$.
Therein, the A-AFM overcomes the ferromagnetic order 
(\Fref{fig:mag phase diagram GCMO})
because the strength of the magnetic coupling
decreases and 
% with respect to referee B13
the effective out-of-plane interaction
becomes negative again (\Fref{fig:jgcmo}).
All three exchange parameters 
are of a similar magnitude, 
$J_x = \SI{1.20}{\meV}$,
$J_z = \SI{0.70}{\meV}$, and 
$J_{xz} = \SI{-0.47}{\meV}$.
This 
energetic 
competition reduces also the total energy of other
magnetic structures and makes them more likely. 
The smallest
energy difference is realized by the FiM state
(\Fref{fig:mag phase diagram GCMO}a) but also G-AFM
and C-AFM show very small energy differences and
might become more relevant.
In particular, the C-type AFM order is also assumed for $x = 0.8$ 
by Beiranvand \etal \cite{BEIRANVAND2017126} but remains
in our calculation  at $x = \nicefrac{7}{8}$ still
\SI{11}{\meV} higher in energy than the A-AFM. 

This variation of potential antiferromagnetic structures
offers a large playground for the study of 
basic principles in magnetic coupling and the 
resulting ground states. Hence, the electron doping 
concentration range
  $\nicefrac{7}{8} \leq x < 1$
is,
in particular, scientifically interesting, because
the experimental results vary a lot:
Beiranvand \etal \cite{BEIRANVAND2017126}
did not detect an CO state for $ x > 0.7$, but
Khan \etal \cite{KHAN2015401} found that
it should coexist with OO simultaneously at $x = 0.85$ 
and be even very robust against external influence, since 
the application of a magnetic field up
to \SI{15}{\tesla} between \SIrange{5}{300}{\kelvin} did 
not annihilate the charge ordering.
In addition, colossal magneto resistance 
was detected at $0.8 < x < 0.9 $ and $T=\SI{10}{\kelvin}$,
in the boundary between the CO-AFM insulating state and the
cluster-glass (CG) state \cite{BEIRANVAND2017126}.
The latter was explained by the simultaneous existence of FM metallicity
and an AFM insulating state \cite{PhysRevB.62.6442}.

%We conclude that further investigations
%-- including both theoretical and experimental techniques --  
%of this GCMO doping region are certainly required.

% ..........................................................
% SUMMARY ..................................................
% ..........................................................
\section{Summary} % ........................................
\label{sec:summary}

We investigated theoretically the magnetic phase diagram of the 
whole GCMO series for the first time and observed a good qualitative 
agreement with the available experimental data \cite{BEIRANVAND2017126}.
We identified the different magnetic ground states 
being mainly a ferromagnetic coupling between the Mn magnetic moments
with instabilities towards ferrimagnetic or A-type 
antiferromagnetic spin orientations. The calculated magnetic transition 
temperatures agree well with the experimentally derived 
ones but show a systematic difference to experiment for $x>\SI{60}{\percent}$. 
This might be connected with the 
unstable antiferromagnetic coupling between the Mn ions observed in the 
same concentration range.  In summary, we obtained a rather good agreement
between the numerical calculations based on the 
special quasi random structures simulating the miscibility of 
the GCMO series and the earlier experimental study of the whole 
concentration range \cite{BEIRANVAND2017126}. 
	%referee B15
Several concepts remain still unknown for GCMO and need to be carefully 
examined e.g.: Does GCMO favor a collinear or non-collinear magnetism?
Which combination of spin, charge and orbital ordering is likely to occur in 
GCMO? What is the effect of strain or defects on the magnetic phase diagram. 
Thereby, our	study lays a basis for further experimental and theoretical 
studies of the solid solution rare earth manganites and in particular GCMO.
% ..........................................................
% ..........................................................
% ..........................................................
\section*{Acknowledgments} % ...............................
This publication was funded by the German Research Foundation
within the Collaborative Research Centre 762 
(Projects No. A4 and No. B1) and the 
German Academic Research Council (Project number 57348127). 
The Jenny and Antti Wihuri
Foundation is acknowledged for financial support.
The computer resources of the Finnish IT Centre for
Science (CSC), project No. 2000643, are acknowledged.
% ..........................................................
% ..........................................................
% ..........................................................
% APPENDIX
%\appendix
%\section{Supplemental material}

%%%%%%%%%%%%%%%%%%%%%%%%%%%%%%%%%%%%%%%%%%%%%%%%%%%%%%%%%%%%
% BIBLIOGRAPHY .............................................
\bibliographystyle{apsrev4-1}
\bibliography{journals,references}
%%%%%%%%%%%%%%%%%%%%%%%%%%%%%%%%%%%%%%%%%%%%%%%%%%%%%%%%%%%%
\end{document}